\documentclass[preprint,showpacs,aps,preprintnumbers,amsmath,amssymb]{revtex4-1}

\usepackage{graphicx,epsfig}
\usepackage{dcolumn}
\usepackage{bm}
\newcommand{\be}{\begin{equation}}
\newcommand{\ee}{\end{equation}}
\newcommand{\bse}{\begin{subequations}}
\newcommand{\ese}{\end{subequations}}
\newcommand{\bary}{\begin{eqnarray}}
\newcommand{\eary}{\end{eqnarray}}
\newcommand{\bpmat}{\begin{pmatrix}}
\newcommand{\epmat}{\end{pmatrix}}
\newcommand{\bwt}{\begin{widetext}}
\newcommand{\ewt}{\end{widetext}}

\begin{document}


\title{Propagation of GeV neutrinos through Earth}
\author{Yaithd Daniel  Olivas$^{a}$}
\email{yaithd.olivas@correo.nucleares.unam.mx}
\author{Sarira Sahu$^{a,b}$ }
\email{sarira@nucleares.unam.mx}

\affiliation{$^{a}$Instituto de Ciencias Nucleares, Universidad Nacional Aut\'onoma de M\'exico, 
Circuito Exterior, C.U., A. Postal 70-543, 04510 Mexico DF, Mexico}


\affiliation{$^{b}$Astrophysical Big Bang Laboratory, RIKEN, Hirosawa, Wako, Saitama 351-0198, Japan}


\begin{abstract}
We have studied the Earth matter effect on the oscillation of upward
going  GeV
neutrinos by taking into account the three active neutrino
flavors. For neutrino energy in the range 3 to 12 GeV we observed
three distinct resonant peaks for the oscillation process
$\nu_e\leftrightarrow \nu_{\mu,\tau}$ in three \textit{distinct}
densities. However, according to the most realistic density profile of
the Earth, the second peak at neutrino energy 6.18 GeV corresponding
to the density $6.6\,g/cm^3$ does not exist. So the resonance at this
energy can not be of MSW-type. For the calculation of  observed flux
of these GeV neutrinos on Earth, we considered two different flux
ratios at the source, the standard scenario with the flux ratio $1:2:0$ and
the muon damped scenario with $0:1:0$. It is observed that at the
detector while the standard
scenario gives the observed flux ratio $1:1:1$, the muon damped
scenario has a different ratio. For muon damped case with $E_{\nu} < 20$ GeV, 
we always get observed neutrino fluxes as $\Phi_{\nu_e} <
\Phi_{\nu_\mu}\simeq \Phi_{\nu_\tau}$ and for $E_{\nu} > 20$ GeV, we
get the average
$\Phi_{\nu_e}\sim 0$ and $\Phi_{\nu_\mu}\simeq \Phi_{\nu_\tau}\simeq 0.45$.
The upcoming PINGU will be able to shed more
light on the nature of the resonance in these GeV neutrinos and
hopefully will also be able to discriminate among different processes of neutrino
production at the source in GeV energy range.

\end{abstract}

\maketitle

\section{Introduction}
\label{intro}

During the last couple of decades, a significant amount of information about the neutrino properties have
been obtained by many experiments\cite{Fukuda:1998mi,Ashie:2005ik,Smy:2003jf,Gando:2010aa,An:2013uza} and now neutrino physics has entered an era of
precision measurement and deeper understanding of the oscillation
phenomena. The recent observation of TeV-PeV neutrino events by
IceCube in South Pole for the first time shows the cosmological origin
of these high energy neutrinos\cite{Aartsen:2013jdh,Aartsen:2014gkd}, although the sources and the
production mechanism  are still unknown. The DeepCore subarray\cite{Collaboration:2011ym} of the
IceCube has the energy threshold of about 10 GeV which can study low
energy neutrino physics. Also below 100 GeV the DeepCore increases the
effective area of IceCube by more than an order of magnitude. So the
DeepCore subarray has opened up a new window on GeV neutrino oscillation
physics, mostly the atmospheric neutrino oscillation. The next
generation upgrade to IceCube is the Precision IceCube Next Generation
Upgrade (PINGU) \cite{Palczewski:2014mza}. This will deploy an additional 20 strings within the
DeepCore to lower the sensitivity from ${\cal O}(10)$  GeV to ${\cal O}(1)$ GeV. The
goal of PINGU is to perform precise measurements of
atmospheric neutrino oscillations down to a few GeV and to determine
the neutrino mass hierarchy.

The matter effect on the neutrino oscillations is being studied in
different context\cite{Kuo:1986sk,Smirnov:1987mk,Petcov:1986qg,Walker:1986xd,Jegerlehner:1996kx,Ohlsson:1999um,Ohlsson:1999xb,Ohlsson:2001et,Ohlsson:2001ck,Ohlsson:2001vp,Sanroma:2011uj,Dighe:2003vm,Akhmedov:2008qt,Akhmedov:2006hb,Rott:2015kwa}. The neutrino properties get modified due to the
medium effect. Even a massless neutrino acquires an effective mass and
an effective potential in the matter. When the neutrinos from the
interior of the sun propagate out, they can undergo resonant
conversion from one flavor to another due to the medium effect which
is well known as the Mikheyev-Smirnov-Wolfenstein (MSW)
effect\cite{Wolfenstein:1977ue,Mikheev:1986gs}. Similarly, the
neutrino propagation in the supernova medium\cite{Walker:1986xd,Fuller:1998kb,Takahashi:2002yj,Yoshida:2006qz,Dasgupta:2007ws,Duan:2010bf,Sahu:1998jh},
in Gamma-Ray Burst (GRB) fireball\cite{Meszaros:2001ms,Razzaque:2002kb,Razzaque:2003uv,Kajino:2008zz,Sahu:2005zh,Sahu:2009iy}, in Choked GRBs\cite{Razzaque:2005bh,Razzaque:2004yv,Mena:2006eq,Razzaque:2014vba,Razzaque:2009kq,Senno:2017vtd,Murase:2013ffa,Sahu:2010ap,Oliveros:2013apa} and early universe
hot plasma\cite{Enqvist:1990ad} can have many important implications in their respective
physics. The neutrino propagation in the Earth has also been studied
in various context and different approximations to the Earth density
profile are considered \cite{Ohlsson:1999um,Ohlsson:2001et,Ohlsson:2001ck,Freund:1999vc,Winter:2005we,Agarwalla:2012uj,Winter:2015zwx}. For most of the realistic calculations, the Preliminary
Reference Earth Model (PREM)\cite{PREM} density profile is
considered. In this case the density obtained is function of depth from
the surface of the Earth and both longitudinal and latitudinal
variations are ignored. Also  the density profile of the Earth is
symmetric on both sides of the centre.
All oscillation experiments of same
baseline length will have the same matter effect.

In the energy range of 1 to
100 GeV, the atmospheric neutrinos are the largest contributor to the
background in the detector and it has been studied in
detail\cite{Winter:2005we,Agarwalla:2012uj,Winter:2015zwx}. Detection of any astrophysical neutrinos in this energy range
is difficult due to the overwhelming atmospheric neutrino
background. While the DeepCore increases the effective area of IceCube
by one order of magnitude for neutrino energy below 100 GeV, the next
generation IceCube upgrade PINGU has low sensitivity $\sim {\cal O}(1)$ GeV and will be able to detect these
low energy neutrinos. Our aim here is to study these neutrinos in the
energy range $1 \, GeV \le E_{\nu} \le 100\, GeV$ of
astrophysical origin. There are many astrophysical transient objects e.g. 
GRBs\cite{Gao:2011xu,Bartos:2013hf,Murase:2013hh,Meszaros:2000fs,Bahcall:2000sa,Murase:2013mpa},
and AGN\cite{Atoyan:2004pb} which are
potential sources of these neutrinos. Detection of these neutrinos
in spatial and temporal correlation with the gamma-rays/X-rays from
these GRBs and flaring blazars (blazar is a subclass of AGN) is
possible. Detection of these neutrinos will be important to understand
the production and acceleration mechanisms in these sources. Also, these low
energy neutrinos can have resonant oscillation within the
Earth which is absent for higher energy neutrinos. So this inspires us
to study the matter effect on the oscillation of multi-GeV
neutrinos when crossing the diameter of the Earth and the modification in
their flux ratios.

Here we would like to consider the oscillation of three flavor neutrinos to study the
Earth matter effect on the upward going neutrinos  and the possible
modification of their flux ratio at the detector.
The paper is organized as follows: In Sec. 2 we discuss about the
formalism used to calculate the neutrino oscillation probability in
the presence of a matter background. A realistic
Earth density profile is discussed in Sec. 3. In Sec. 4 we elaborate on
our results and a comprehensive discussion is given in Sec. 5.

\section{Formalism}
\label{sec:2}

The neutrino oscillation in vacuum and matter has been discussed
extensively for solar, atmospheric, as well as accelerator and
reactor experiments. Models of three active flavor neutrinos oscillation in
constant matter
density\cite{Barger:1980tf,Kim:1986vg,Zaglauer:1988gz}, 
linearly varying density and exponentially varying density have been 
studied\cite{Petcov:1987xd,Lehmann:2000ey}. In Ref.\cite{Ohlsson:1999xb}, 
T. Ohlsson and H. Snellman have developed an analytic formalism for
the oscillation of three flavor neutrinos in the matter background
with varying density, where they use the plane wave approximation for
the neutrinos (henceforth we refer to this as OS formalism). Here the
evolution operator and the transition probabilities are expressed as
functions of the vacuum mass square differences, vacuum mixing
angles and the matter density parameter. As applications of the above
formalism, the authors have studied the neutrino oscillation
traversing the Earth and the Sun for constant, step-function and
varying matter density profiles\cite{Ohlsson:1999um,Ohlsson:2001et}. To handle the varying
density, the distance is divided into equidistance slices and in each
slice the matter density is assumed to be constant. Recently this
formalism is also used to study the multi-TeV neutrino propagation in
the choked GRBs\cite{Oliveros:2013apa} and the calculation of the track to shower ratio of
the multi-TeV neutrinos in IceCube\cite{Varela:2014mma}. In this section we review the OS
formalism for the calculation of neutrino oscillation probability.

In the context of three active neutrino flavors, a flavor neutrino
state can be expressed as a linear superposition of mass eigenstates
as
\be
 |\nu_{\alpha}\rangle = \sum^{3}_{i=1} U^*_{\alpha i}|\nu_{i}\rangle , 
\label{flavormatter}
\ee
where $\alpha = e, \mu, \tau$ (flavor eigenstates) and $i = 1,2,3$ (mass
eigenstates). The $U_{\alpha i}$ is the three by three neutrino mixing
matrix given by,
\bwt
 \bary
   && U = 
    \bpmat
    U_{e1} & U_{e2} & U_{e3}\\
    U_{\mu 1} & U_{\mu 2} & U_{\mu 3}\\
    U_{\tau 1} & U_{\tau 2} & U_{\tau 3}\\
 \epmat
\nonumber\\
  & &=
    \bpmat
    c_{13} c_{12} & c_{13} s_{12} & s_{13} e^{-i\delta_{cp}} \\
  -s_{12} c_{23} - c_{12} s_{23} s_{13} e^{i\delta_{cp}} &  c_{12} c_{23} - s_{12} s_{23} s_{13} e^{i\delta_{cp}}  & s_{23} c_{13}\\
s_{23} s_{12} - c_{23} s_{13} c_{12} e^{i\delta_{cp}} &  - s_{23} c_{12} - s_{13} s_{12} c_{23} e^{i\delta_{cp}} & c_{23}c_{13}\\
 \epmat,
  \label{pmns}
  \eary
%
\ewt
where $c_{ij} \equiv \cos\theta_{ij}$ and $s_{ij} \equiv
\sin\theta_{ij}$ for $i,j = 1,2,3$. 
With three neutrino flavors, there are three neutrino mixing angles
$\theta_{12}, \theta_{13}$, $\theta_{23}$ and CP violating phase
$\delta_{CP}$. In the present analysis we take $\delta_{CP}=0$ since
CP non conservation is negligible at the present level of accuracy
hence the entries of the CKM matrix are real numbers. 

Propagating neutrinos in a medium experience an effective potential
due to the collision with the particles in the background matter.
Depending on the neutrino flavor  the interaction can be  charged current (CC)
or neutral current (NC) or both. The neutral current interaction is same for
all the neutrinos which can be factored out as a global phase and
only charged current term will contribute. This is attributed only to
electron neutrino and its anti-neutrinos. The effective potential is
expressed as 
\be
  \label{eqn:vf2}
V_{f}=A 
  \begin{pmatrix}
    1 & 0 & 0\\
    0 & 0 & 0\\
    0 & 0 & 0
  \end{pmatrix},
\ee
where $A= \pm\sqrt{2}G_{F}N_{e}$, $G_{F}$ is the Fermi coupling
constant and $N_{e}$ represents the electron number density in the
background medium and  signs $\pm$
correspond to $\nu_e$ and ${\bar \nu}_e$ respectively.

In vacuum, the Hamiltonian that described the propagation of the neutrinos in the mass 
eigenstate basis  is described by 
\be
H_{m} = 
\bpmat
E_{1} & 0 & 0 \\
0 & E_{2} & 0 \\
0 & 0 & E_{3} \\
\epmat,
\label{Hm}
\ee
where $E_{i}$, for $i=1,2,3$ refer to the energy of each neutrino mass
eigenstate with
\be
E_i=\sqrt{{\bf p}^2+m^2_i}.
\ee 
Here we assume that neutrinos with different masses have the same momentum.
This Hamiltonian can be 
written in the flavor basis through the unitary transformation
described by the matrix $U$ from equation \eqref{pmns}, as
\be
 H_{f} = UH_{m}U^{-1}.
\ee
In the mass basis, the total Hamiltonian is given by
\bary
 {\cal H}_{m} &=& H_{m} + U^{-1} V_{f} U\\ \nonumber
                    &=& H_{m}+V_{m}.
 \label{htot}
\eary
The total Hamiltonian in the flavor basis is written as
\be
{\cal H}_{f}=H_f + V_f.
\ee
For neutrino propagation in a medium, the Hamiltonian is not diagonal,
neither in the mass basis nor in the flavor basis, so one has to
calculate the evolution operator in any of these basis.
In the mass basis, the evolution of the state at a later time $t$ will be obtained by solving the
Schr\"{o}dringer  equation 
\be
 i\frac{d |\nu_{i} (t) \rangle}{dt} = {\cal H}_{m} |\nu_{i} (t)\rangle,
\ee
 and the solution to this equation can be expressed in terms of the evolution operator as
\bary
 |\nu_{i} (t)\rangle &=& e^ {-i {\cal H}_{m} t} |\nu_{i} (0)\rangle \\ \nonumber 
                    &=& U_{m}(t) |\nu_{i} (0)\rangle,
 \label{evol}
\eary
where $U_{m}(t)=e^{-i\,{\cal H}_m t}$ is the evolution operator in the mass basis and in 
the flavor basis this can be written as
\be
U_{f}(t) = U U_{m}(t) U^{-1}.
\ee
Neutrinos being relativistic, we can replace $t$ by the path length $L$,
where we use the natural units $c=1$ and $\hbar=1$ . 

The evolution operator of Eq.(\ref{evol}) can be computed using the
definition of the exponential of a matrix but it is not a
straightforward task since the definition implies an infinite sum. The
Cayley-Hamilton theorem provides a powerful tool to reduce this
infinite sum to a finite sum and is given by
\be
  \label{eqn:exp}
  \begin{aligned}
  e^{-i{\cal H}_{m}t}&=e^{-iTt-\frac{i}{3}(Tr{\cal H}_{m})It}\\
  &=\phi e^{-iTt}\\
&=\phi\left[a_{0}I+a_{1}(-iTt)+a_{2}(-iTt)^2\right]\\
& = \phi\left(a_{0}I-ia_{1}tT-a_{2}t^{2}T^2\right),
\end{aligned}
\ee
where we define the traceless matrix $T={\cal H}_{m}-\frac{1}{3}Tr({\cal H}_{m})\, I$ and $I$ is
the identity matrix. The final
expression for the evolution operator is given by (by replacing $t$ to
$L$) 
\be
  e^{-i{\cal H}_{m}L}=\phi\left(a_{0}I-ia_{1}LT-a_{2}L^{2}T^2\right).
\label{cheq}
\ee
In order to determine the evolution operator it is necessary to know
the coefficients $a_i$ in Eq.(\ref{cheq}). The $T$ matrix has three
eigenvalues $\lambda_i$ with $i=1,2,3$ and the characteristic equation
is
\be
\lambda^3+c_2\lambda^2+c_1\lambda+c_0=0.
\label{chareq}
\ee
The coefficients of $\lambda$ are given as
\be
c_0=-{\rm det}(T),\,\, c_2=-tr(T)=0, 
\ee
and
\be
c_1=T_{11}T_{22}-T^2_{12}+T_{11}T_{33}-T^2_{13}+T_{22}T_{33}-T^2_{23}.
\ee
This reduces the Eq.(\ref{chareq}) to 
\be
\lambda^3+c_1\lambda+c_0=0,
\label{chareq}
\ee
and the eigenvalues are given as
\bary
\lambda_1&=& \frac{X}{2^{1/3} 3^{2/3}} -\frac{\left (\frac{2}{3}\right
)^{1/3}c_1}{X},\nonumber\\
\lambda_{2,3}&=& \frac{(1\pm i\sqrt{3}) c_1}{2^{2/3} 3^{1/3} X}
-\frac{(1\mp i\sqrt{3}) X}{2\times 2^{1/3} 3^{2/3}},
\eary
with
\be
X=\left (  
\sqrt{3} \sqrt{ 4 c_1^3 + 27 c_0^2}-9c_0 
\right )^{1/3}.
\ee
With the use of the above equations, the evolution operator in the mass basis can
be written as
\bary
U_{m}(L) &=& e^{-i {\cal H}_{m} L} \\ \nonumber
         &=& \phi \sum^{3}_{a=1} e^{-iL\lambda_{a}} \frac{\left[ (\lambda^{2}_{a} + c_{1}) I + \lambda_{a} T + T^{2} \right]}{3
           \lambda^{2}_{a} 
+ c_{1}},
\label{uevol2}
\eary
The evolution operator in the flavor basis 
is given by 
\bary
U_{f} (L) &=& e^{-i {\cal H}_{f} L} \\ \nonumber
         &=& U e^{-i {\cal H}_{m} L} U^{-1} \\ \nonumber
         &=& \phi \sum^{3}_{a=1} e^{-iL\lambda_{a}} \frac{\left[ (\lambda^{2}_{a} + c_{1}) I + \lambda_{a} \tilde{T} + \tilde{T}^{2} \right] }{3 \lambda^{2}_{a} + c_{1}},
 \label{evolflavor}
\eary
where $\tilde{T}$ is in the flavor basis.

The probability 
of flavor change from a flavor $\alpha$ to another flavor $\beta$ due to neutrino oscillation
through a distance L can be given by
\bary
&&P_{\nu_{\alpha}\rightarrow{\nu_{\beta}}}(L) \equiv 
 P_{\alpha \beta}(L) = |\langle \nu_{\beta} | U_{f}(L) | \nu_{\alpha}
 \rangle|^{2} \nonumber\\
&=& \delta_{\alpha \beta} - 4 \underset { a < b}{\sum^3_{a=1} \sum^3_{b=1}}
P_{a}(L)_{\beta\alpha} P_{b}(L)_{\beta\alpha} \sin^2 x_{ab}, 
 \label{prob}
\eary
where we have defined
\be
P_{a}(L)_{\beta\alpha} =\frac{
(\lambda^2_a+c_1)\delta_{\beta\alpha} +
  \lambda_a {\tilde T}_{\beta\alpha} + {\tilde T^2}_{\beta\alpha}
}{3 \lambda^2_a + c_1}.
\ee
The matrices ${\tilde T}_{\beta\alpha}$ and ${\tilde
  T^2}_{\beta\alpha}$ are symmetric and defined as
\be
{\tilde T}_{\alpha\beta}={\tilde T}_{\beta\alpha}=\sum^3_{a=1} \sum^3_{b=1} U_{\alpha a}
U_{\beta b} T_{ab},
\ee 
and
\be
{\tilde T^2}_{\alpha\beta}={\tilde T^2}_{\beta\alpha}=\sum^3_{a=1} \sum^3_{b=1} U_{\alpha a}
U_{\beta b} T^2_{ab}.
\ee
Also we have defined the quantity
\be
x_{ab}=\frac{ (\lambda_a-\lambda_b) L}{2}.
\ee
The matrix T is written explicitly as 
\be
T_{ab}= 
\bpmat
T_{11}
&
A U_{e1} U_{e2} & A U_{e1} U_{e3} \\
A U_{e1} U_{e2} & 
T_{22} & A U_{e2} U_{e3}
\\
A U_{e1} U_{e3} & A U_{e2} U_{e3}  & 
T_{33}
\\
\epmat,
\label{tmatrix}
\ee
where the diagonal elements of the above matrix are given by
\be
T_{aa}=A U^2_{ea} + \frac{1}{3} (\sum^3_{b\neq a=1}E_{ab}-A).
\ee
Here $E_{ab}=-E_{ba}=E_a - E_b$ and
the energies satisfy the relation 
\be
E_{12}+E_{23}+E_{31}=0.
\ee
The neutrino oscillation probabilities satisfy the condition
\be
\sum_{\beta} P_{\alpha\beta}=1,\,\,\, {\rm for}\,\,\,
\alpha,\beta=e,\mu,\tau, 
\label{probsum}
\ee
and a similar condition is satisfied for anti-neutrinos which we
define as $P_{{\bar\alpha}{\bar\beta}}$.

Using the Eqs.(\ref{prob}) and (\ref{probsum}) we can calculate the probability of
transition from one flavor to another.  For $V_f=0$,  we get the
vacuum transition probability. For matter with varying
density the distance $L$ can be discretized into small intervals
$[L_i,\, L_{i+1}]$ in such a way that the density profile is almost
constant in each segment and can be used this procedure repeatedly in
each segment.
By doing so we can study numerically the neutrino
oscillation in any type of density profile. For neutrinos traversing a
series of matter densities $\rho_i$ for $i=$ 1 to $n$, with their
corresponding thickness $L_i$, the total evolution operator is the
ordered product and is given as
\be
U_f(L)=\prod^n_i U_f(L_i),
\label{produf}
\ee 
where $\sum^n_i L_i=L$.
In a series of papers by OS, this method has been applied for
different density profiles of the Sun and the Earth, to study the MeV energy
neutrino
oscillation\cite{Ohlsson:1999xb,Ohlsson:2001et,Ohlsson:1999um,Freund:1999vc}. 
To check the consistency of our numerical method, we used the constant
matter density, mantle-core-mantle step function as well as realistic
Earth matter density profiles and reproduced the results of Figs. 2, 3
and 4 of ref. \cite{Freund:1999vc}. We also reproduced the results obtained in
Fig.1 to Fig. 6 of ref. \cite{Ohlsson:1999xb} to establish the
correctness of our numerical method.

\section{Earth Density Profile}

\begin{figure}[t!]
{\centering
\resizebox*{0.55\textwidth}{0.38\textheight}
{\includegraphics{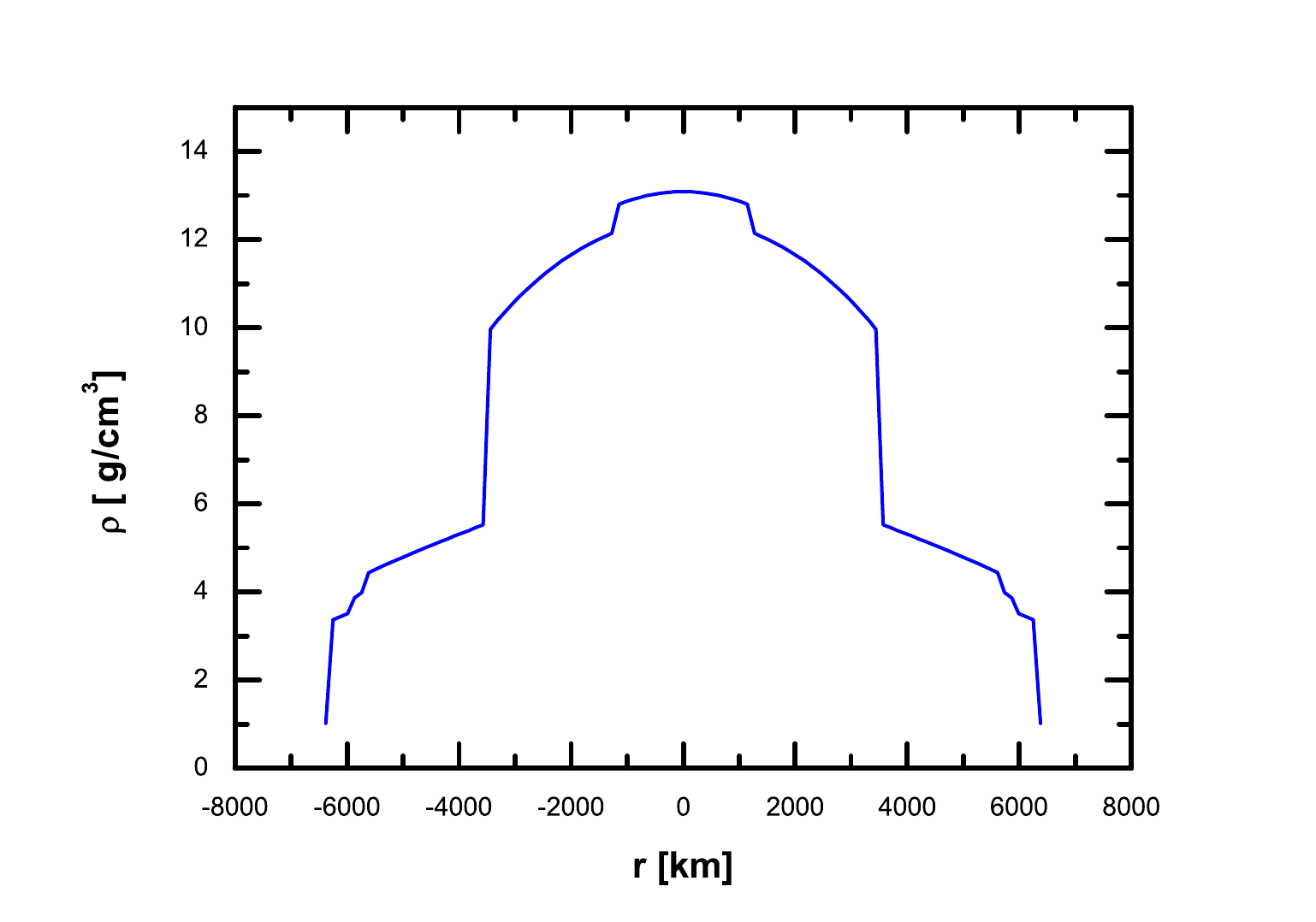}}
\par}
\caption{
The Earth density profile PREM is plotted as a function
    radius $r$.
}
\label{fig:density}
\end{figure}

 High energy neutrinos reaching the detector like IceCube from opposite
side of the Earth can experience both oscillation and  absorption due to
CC and NC interactions. While the oscillation is important for low energy neutrinos $E_{\nu} \le 10$ TeV, 
for very high energy neutrinos the interaction
cross sections are large  so that the absorption effects become
very important and have to be taken into account as the shadowing
effect\cite{Varela:2014mma}. But here we are
considering the multi-GeV neutrinos, so the absorption effect is very
small and we don't take into account.
Although, the density profile of the Earth is not known
exactly, here we consider the the most realistic density profile
Preliminary Reference Earth Model (PREM)\cite{PREM}
which is given as 
\bwt 
\begin{eqnarray}
  \label{earthdensity}
  \rho(x)=
  \begin{cases}
    13.0885-8.8381x^2, \qquad \qquad \qquad \qquad \qquad \qquad 0\leq x\leq 0.191\\
    12.5815-1.2638x-3.6426x^2-5.5281x^3, \qquad 0.191<  x\leq 0.546\\ 
    7.9565-6.4761x+5.5283x^2-3.0807x^3, \qquad 0.546<  x\leq 0.895\\ 
    5.3197-1.4836x, \qquad \qquad \qquad \qquad \qquad \qquad 0.895<  x\leq 0.905\\ 
    11.2494-8.0298x, \qquad \qquad \qquad \qquad \qquad \qquad 0.905<  x\leq 0.937\\ 
    7.1089-3.8045x, \qquad \qquad \qquad \qquad \qquad \qquad 0.937<  x\leq 0.965\\
    2.6910+0.6924x, \qquad \qquad \qquad \qquad \qquad \qquad 0.965<  x\leq 0.996\\
    2.900, \qquad \qquad \qquad \qquad \qquad \qquad \qquad \qquad \quad 0.996<  x\leq 0.997\\
    2.600, \qquad \qquad \qquad \qquad \qquad \qquad \qquad \qquad \quad 0.997<  x\leq 0.999\\
    1.020,\qquad \qquad \qquad \qquad \qquad \qquad \qquad \qquad \quad 0.999<  x\leq 1.\\
  \end{cases}
\end{eqnarray}
\ewt
Here $x=r/R_{\oplus}$, $R_{\oplus}=6367$ km is the radius of the Earth
and the density $\rho$ is in units of
${\text g/cm^3}$ which is shown in Fig. \ref{fig:density} as a function of $r$. The density
profile is symmetric around the centre of the Earth and independent of
the longitudinal and latitudinal variations.

\section{Results}

In the standard picture of neutrino oscillation, the oscillation
experiments with solar, atmospheric, reactor and accelerator neutrinos
can be explained through the parameters\cite{An:2013uza,Ashie:2004mr,Araki:2004mb} 
\bary
  \Delta m^2_{21}=8.0\times 10^{-5} \mathrm{eV}^2, \quad
  \theta_{12}=33.8^{\circ}, \quad \quad \theta_{23}=45^{\circ} \nonumber \\
  \Delta m^2_{31}=3.2\times 10^{-3} \mathrm{eV}^2, \quad
  \theta_{13}=8.8^{\circ} \quad \text {and} \quad \delta_{CP}=0,
  \eary
with $\Delta m^2_{ij}=m^2_i-m^2_j$. Throughout our analysis we will be
using the above neutrino parameters and
the neutrinos in the energy range $1 \, GeV \le E_{\nu} \le 100\,
GeV$. Also we consider the normal neutrino mass hierarchy 
i.e. $m_e < m_{\mu} < m_{\tau}$ 
for the calculation
of the oscillation probabilities of different neutrino flavors. 
The neutrinos propagating through the Earth will follow different
trajectories depending on the zenith angle $\theta_z$ and is defined
in ref.\cite{Petcov:1998su,Chizhov:1999az}, where $\theta_z=180^{\circ}$
corresponds to vertically
up going neutrinos by crossing the diameter of the Earth, and $\theta_z=90^{\circ}$
corresponds to neutrinos coming from the horizon. To have 
maximum matter effect, we  only consider the neutrinos which have
$cos\theta_z=-1$ so that neutrinos can cross both mantle and core.

\begin{figure}[t!]
{\centering
\resizebox*{0.55\textwidth}{0.38\textheight}
{\includegraphics{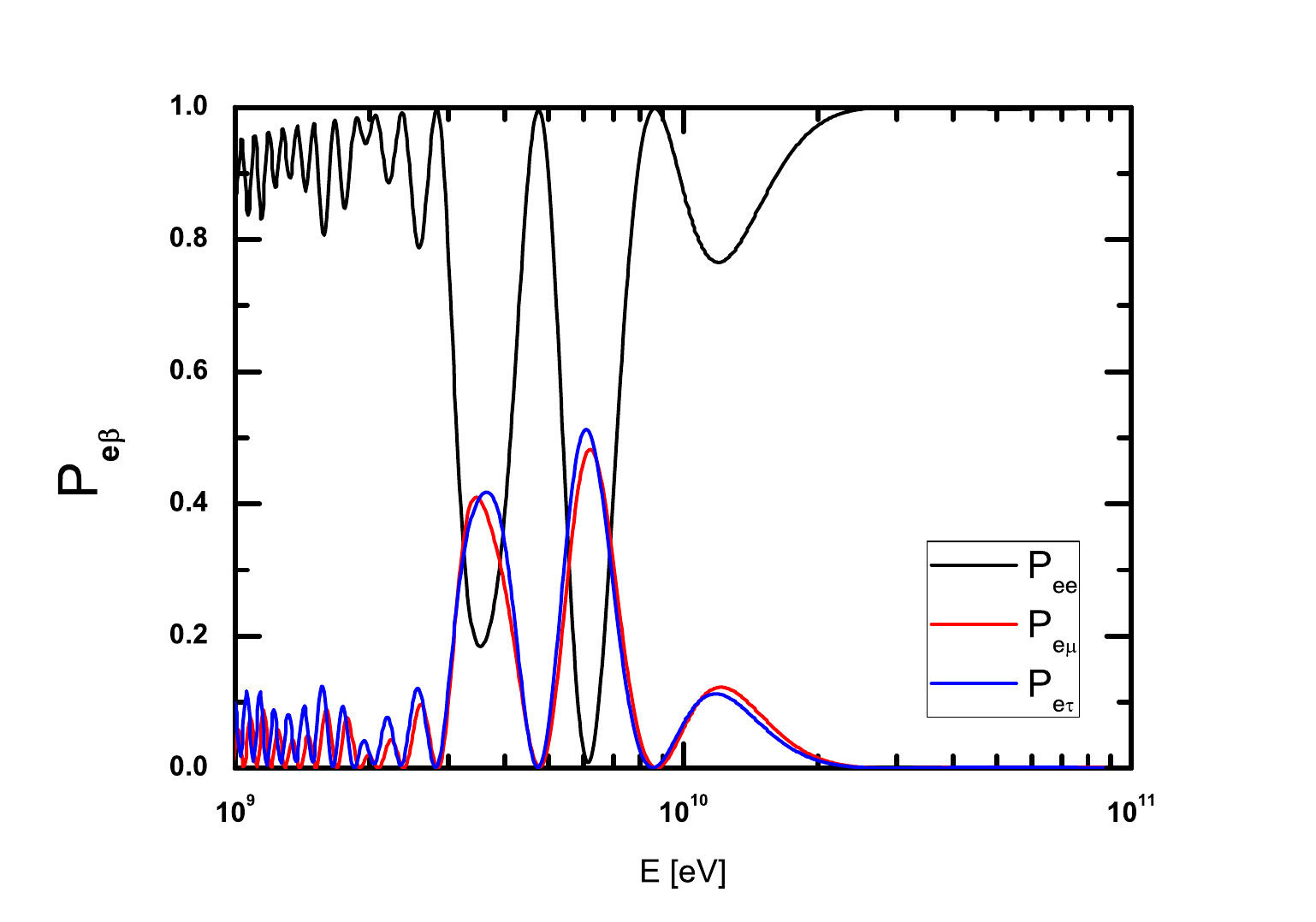}}
\par}
\caption{ $P_{e\beta}$ as a function of neutrino energy $E_{\nu}$.
}
\label{fig:nuenux}
\end{figure}
\begin{figure}[t!]
{\centering
\resizebox*{0.55\textwidth}{0.38\textheight}
{\includegraphics{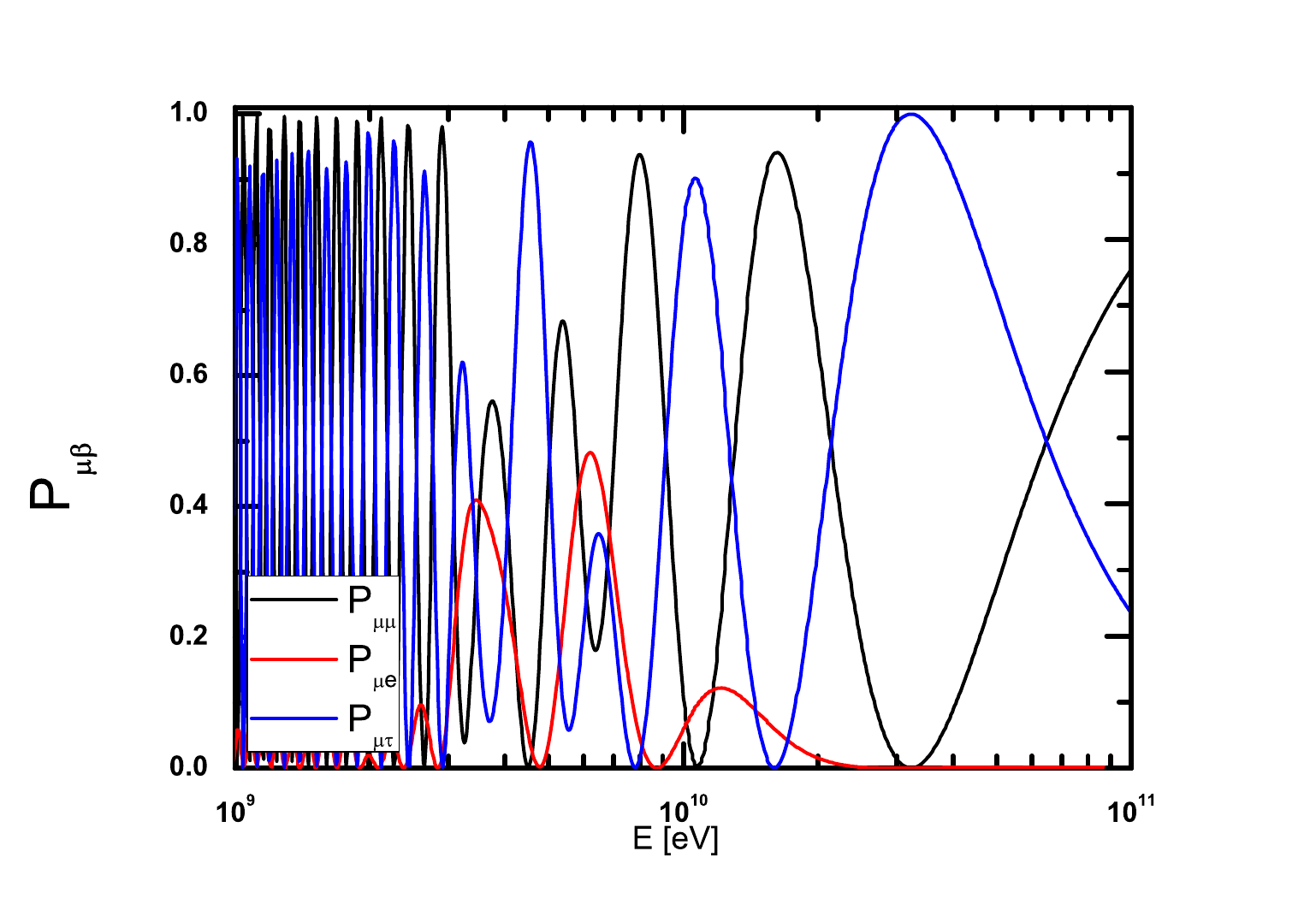}}
\par}
\caption{ $P_{e\beta}$ as a function of neutrino energy $E_{\nu}$.
}
\label{fig:numunux}
\end{figure}

For $\nu_e$ oscillating to $\nu_{\mu}$ and $\nu_{\tau}$ we can
clearly see three distinct resonance peaks in three different neutrino
energies $E_{\nu}$ corresponding to three different densities $\rho$ of
the Earth. The oscillation probability $P_{e\beta}$ is shown in Fig. 2.
For $E_{\nu}=3.45$ GeV, the resonance takes place deep in the core where
$\rho =11.5\, gm/cm^3$ at a depth of $\sim 2483$ km. The third peak
is for $E_{\nu}=12.0$ GeV and the corresponding resonance density and
the distance are respectively $3.4= gm/cm^3$ and 6112 km. These two
peaks are clearly of MSW type because the resonance density and the
resonance length exist for these neutrinos. On the other hand, 
for the second peak with $E_{\nu}=6.18$ GeV, the resonance density 
$\rho =6.6\, gm/cm^3$ does not exist in the Earth's interior (Fig. \ref{fig:density}).
So this resonance can't be of MSW type. This type of resonance is
called parametric resonance which takes place if the variation of the
matter density along the neutrino path is correlated in a certain way
with the change of the oscillation phase
\cite{Ermilova:1986,Akhmedov:1987nc,Akhmedov:1998xq,Liu:1998nb,Akhmedov:2005yj}. 
Below the first resonance peak
($E_{\nu} < 3.45$ GeV) the probability is oscillatory in nature.

In Fig. \ref{fig:nuenux} we have shown
the $P_{ee}$, $P_{e\mu}$ and $P_{e\tau}$ for $1\, GeV \le E_{\nu} \le
100\, GeV$. It shows that for both the oscillations $\nu_e\leftrightarrow \nu_{\mu}$ and 
$\nu_e\leftrightarrow \nu_{\tau}$, the resonance peaks are at the same place for a
given $E_{\nu}$. Beyond $\sim 20$ GeV the transition probabilities are very
small which implies that the Earth's matter does not play any
significant role beyond this energy and the oscillation is purely due
to the vacuum effect.

\begin{figure}[t!]
{\centering
\resizebox*{0.55\textwidth}{0.38\textheight}
{\includegraphics{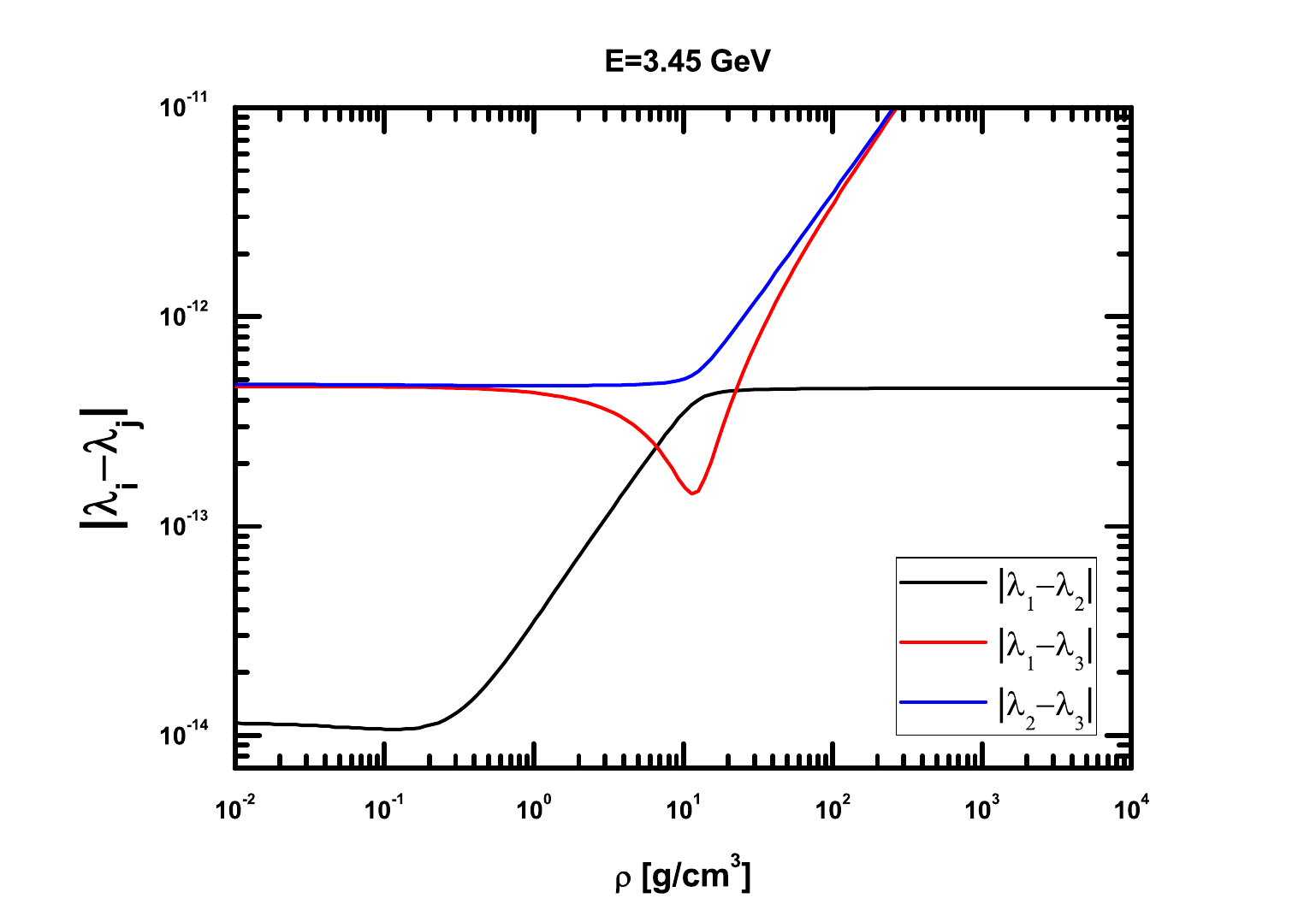}}
\par}
\caption{ Neutrino energy difference is plotted as a function of density
}
\label{fig:lambda1}
\end{figure}

In Fig. \ref{fig:numunux} we have shown the $P_{\mu\mu}$ and transition probabilities of
muon neutrino to $\nu_e$ ($P_{\mu e}$)   and  to $\nu_{\tau}$
($P_{\mu\tau}$). The three resonance peaks in $P_{\mu e}$ are clearly
seen (red curve) but there is no resonant oscillation for
$\nu_{\mu}\leftrightarrow\nu_{\tau}$ (blue curve).
Above about $\sim 20$ GeV  the $P_{\mu e}$ goes to zero and both
$P_{\mu\mu}$ and $P_{\mu\tau}$ are out of phase by
$180^{\circ}$. We observed that for $E_{\nu} > 100$ GeV the $\nu_{\mu}$ does not
oscillate to $\nu_{\tau}$ any more. The oscillation process
$\nu_{\tau}\leftrightarrow\nu_{e}$ is same as $\nu_e\leftrightarrow
\nu_{\mu}$, hence we do not discuss about it.

\begin{figure}[t!]
{\centering
\resizebox*{0.55\textwidth}{0.38\textheight}
{\includegraphics{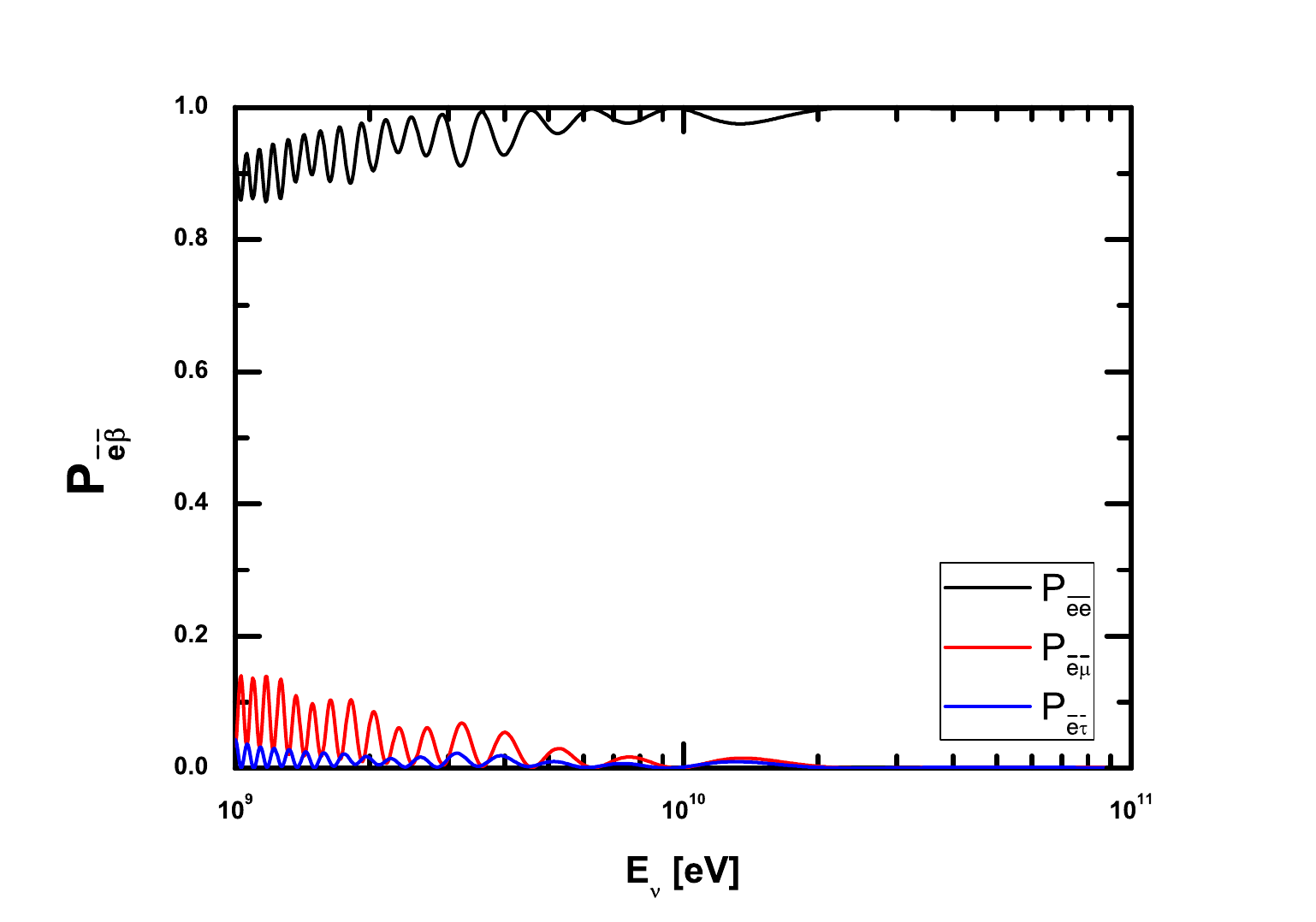}}
\par}
\caption{ $P_{{\bar \mu}{\bar \beta}}$ as a function of $E_{\nu}$.
}
 \label{fig:antinumunux}
\end{figure}

Due to the matter effect the energy eigenvalues of the neutrinos are
given by $\lambda_a$ and the energy difference is related to the effective mass
square difference as
\be
|\lambda_i -\lambda_j | = \frac{|\Delta {\tilde m}^2_{ij}|}{2 E_{\nu}}.
\ee
In Fig. \ref{fig:lambda1} 
we have plotted $|\lambda_i -\lambda_j |$ as a function of Earth density $\rho$ for
resonance neutrino energy $E_{\nu}=3.45$ GeV. It shows that, at the resonance  density $\rho_c$,
both $|\lambda_1 -\lambda_2 |$ and $|\lambda_2 -\lambda_3 |$
have the closest approach and at this point the neutrino mixing is
maximal. Going from the resonance peak at 3.45 GeV to 12 GeV
the resonance density decreases from 11.5 $gm/cm^3$ to
3.4 $gm/cm^3$.

\begin{figure}[t!]
{\centering
\resizebox*{0.55\textwidth}{0.38\textheight}
{\includegraphics{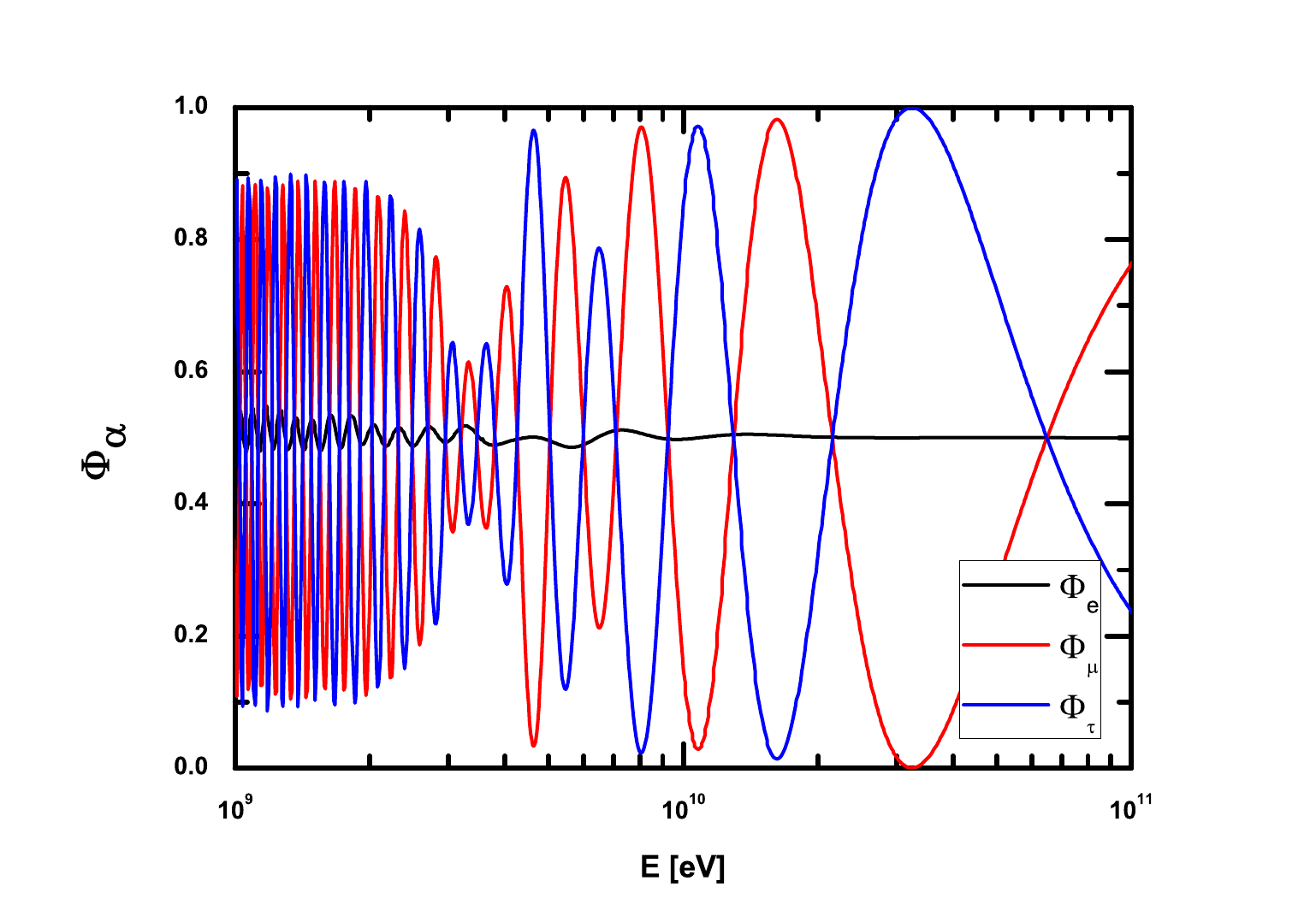}}
\par}
\caption{Neutrino flux at the IceCube detector when $\Phi^0_{\nu_e}:\Phi^0_{\nu_\mu}:\Phi^0_{\nu_\tau}=1:2:0$.
}
\label{fig:flux120}
\end{figure}

\begin{figure}[t!]
{\centering
\resizebox*{0.55\textwidth}{0.38\textheight}
{\includegraphics{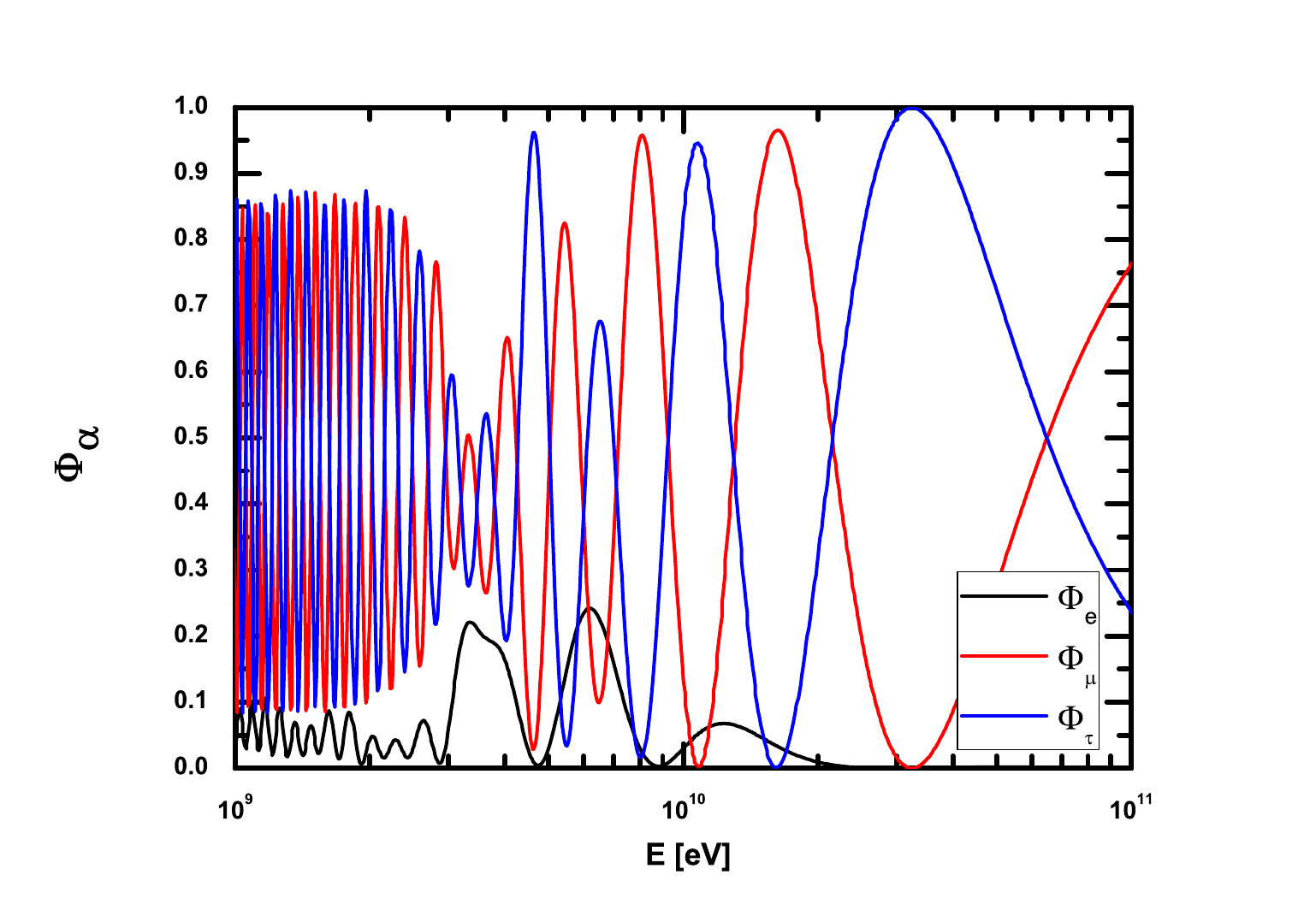}}
\par}
\caption{Neutrino flux at the IceCube detector when $\Phi^0_{\nu_e}:\Phi^0_{\nu_\mu}:\Phi^0_{\nu_\tau}=0:1:0$.
}
\label{fig:flux010}
\end{figure}

There are many uncertainties in determining the density profile of the
Earth. The neutrino oscillations are not very sensitive to
structures and gradients in the density profile if the length scale of
these structures are shorter than the oscillation length\cite{Winter:2015zwx}. However, we
can treat the fluctuation in the density profile PREM by varying it
around the mean value. 
By varying $\pm 10\%$ in the PREM density profile,
Agarwalla et al.\cite{Agarwalla:2012uj} have calculated the transition probability
of $\nu_e\rightarrow \nu_{\mu}$ and survival probability of the
process $\nu_{\mu}\rightarrow \nu_{\mu}$. It is shown that, in the
sensitive region to density variation, the transition probability
$P_{e\mu}$ is enhanced and $P_{\mu\mu}$ is reduced, and also the maxima
and minima are shifted with respect to the negligible matter
effects. As a consequence, the change in the density profile will
shift the resonance position, thus modifying the number of events in
different angular and energy bins. In our case also by varying the
density we expect a similar behavior in resonance position which subsequently
will change transition and survival probabilities of different
neutrinos, but due to the fluctuation in the density, the average of these individual probabilities will not
be very much different from their mean values. So, as a first approximation we consider the PREM
without considering any fluctuation into it to understand the neutrino
propagation and the resonance conditions in the few GeV energy
regime.

We have also done the analysis for the oscillation of anti-neutrinos which are shown
in Fig. \ref{fig:antinumunux}. We observed that there is no resonant oscillation of
the anti-neutrinos. As for anti-neutrinos the potential changes sign,
it will never satisfy the resonance condition. However, if we consider
the inverted mass hierarchy then due to the sign change we can have
resonance for anti-neutrino oscillation but not for neutrino
oscillation. In the low energy limit (between 1 to 10 GeV)
for ${\bar\nu}_e\leftrightarrow {\bar\nu}_{\mu,\tau}$ oscillation, 
the oscillation of ${\bar\nu}_e$ to ${\bar\nu}_{\mu}$ is more
preferable than to ${\bar\nu}_{\tau}$ which can be clearly seen from
Fig. \ref{fig:antinumunux}, but the oscillation probability is small. This is happening due
to $\Delta m^2_{31} \gg \Delta m^2_{21}$. Above 10 GeV the oscillation
${\bar\nu}_e\leftrightarrow {\bar\nu}_{\mu,\tau}$ is suppressed.
We observed that above $E_{\nu} > 100$ GeV the ${\bar\nu}_{\mu}$ does not
oscillate to ${\bar\nu}_{\tau}$ any more which is similar to the
neutrino case discussed above.

The GeV energy neutrinos are produced mostly from the pion decay and
have the standard flux ratio at the production point 
$\Phi^0_{\nu_e}:\Phi^0_{\nu_\mu}:\Phi^0_{\nu_\tau}=1:2:0$
($\Phi^0_{\nu_{\alpha}}$ corresponds to the sum of neutrino and
anti-neutrino flux at the source). Also when the muon is damped, 
the flux ratio at the source is $0:1:0$. 
The flux observed at a distance
$L$ from the source is given by
\be
\Phi_{\nu_\alpha}=\sum_{\beta} \Phi^0_{\nu_\beta} P_{\alpha\beta}, \,\,
\alpha,\beta=e,\, \mu, \,\tau.
\label{obsphi}
\ee
When traveling in the Earth, the neutrinos will oscillates and the
probability $P_{\alpha\beta}$ will be different for different flavors
which is shown in Fig. \ref{fig:nuenux}. By using the above two
neutrino flux ratios $1:2:0$ (standard) and $0:1:0$ (muon damped) at the source, we calculate
the normalized observed flux ratio at the IceCube detector for
the upward going neutrinos. For this calculation we don't take into
account the vacuum effect. Here our main aim is to calculate the
observed flux ratios of the neutrinos at the detector for different flux ratios at the
source without taking into account the vacuum oscillation when
traversing the distance between the source and the Earth.

In Fig. \ref{fig:flux120} we observe that for the flux ratio $1:2:0$ at the source, the 
electron neutrino flux $\Phi_{\nu_e}$ is almost constant $\sim 0.5$ and  the
$\Phi_{\nu_\mu}$ and $\Phi_{\nu_\tau} $ oscillate between 0 and 1
averaging out to $0.5$. So for this case the observed flux ratio
is found to be $1:1:1$. In Fig. \ref{fig:flux010} we have shown the
muon damped scenario. For $E_{\nu} > \, 20$ GeV the  $\Phi_{\nu_e}\sim 0$ but average
$\Phi_{\nu_\mu}\simeq\Phi_{\nu_\tau} =0.45$. 
Again for $E_{\nu} < 20$ GeV
there are three peaks in the normalized flux for $\Phi_{\nu_e}$
corresponding to three resonances as discussed before and shown in
Figs. \ref{fig:nuenux} and \ref{fig:numunux}. In this case we
always get $\Phi_{\nu_e} <
\Phi_{\nu_\mu}\simeq \Phi_{\nu_\tau} $. Due to lower sensitivity of
PINGU $\sim {\cal O}(1)$ GeV it can probe the resonance energy region 
$3 \, GeV < E_{\nu}  \le \, 12\, GeV$ very well. 
In the muon damped scenario, any transient source producing neutrinos in the
few GeV energy range will be detected with a suppressed electron neutrino flux
and enhanced muon and tau neutrino fluxes of equal strength if the  Earth density profile is
correct. However, due to the overwhelming atmospheric neutrino background in
this energy range, it will be hard to detect these
neutrinos unless the flux from the source is high.

\section{Discussion}

Apart from atmospheric neutrinos, there are other astrophysical
sources which can produce low energy GeV neutrinos. 
We used the formalism by Ohlsson
and Snellman in a varying potential to calculate the active-active
neutrino oscillation probability numerically by considering three
active neutrino flavors and the realistic density profile PREM of the
Earth. We observed that in the neutrino energy range 
$3 \, GeV \le E_{\nu}  \le \,12\, GeV$ three distinct
resonances were observed in three
different densities. However, the second resonance at an energy
$E_{\nu}=6.18$ GeV corresponding to the density $6.6\,g/cm^3$
does not exit in the Earth interior. So this resonance is of
non-MSW type but a parametric resonance. We also calculated the observed neutrino
flux for these upward going neutrinos for standard scenario and
the muon damped scenario taking into account the normal neutrino mass
hierarchy. For standard scenario we obtained the
observed flux ratio $1:1:1$ whereas for muon damped
scenario we obtained $\Phi_{\nu_e} <\Phi_{\nu_\mu}\simeq\Phi_{\nu_\tau} $ 
for $E_{\nu} < 20$ GeV and above this energy we obtained 
$\Phi_{\nu_e}\sim 0$, $\Phi_{\nu_\mu}\simeq\Phi_{\nu_\tau}=0.45$.
The fluctuation in the density profile PREM can be considered to
calculate the variation in the oscillation probabilities of different
flavors. This change in density will change the position of resonances.
However, as a first approximation, we consider the PREM without any
fluctuation in it.
The PINGU, which has a lower sensitivity  will probably be able
to probe this low energy range and shed more light on the MSW
mechanism and also it can test the correctness of the Earth
density profile PREM. 

\section*{Acknowledgments}
We are thankful to Shigehiro Nagataki for many useful discussions. 
S.S. is a Japan Society for the Promotion of Science (JSPS) invitational fellow.
The work of S. S. is partially supported by DGAPA-UNAM (Mexico) Project
No. IN110815.





\end{document}